\documentclass[12pt]{iopart}

\usepackage{iopams}
\usepackage{graphicx}
\usepackage{color}

\newcommand{\comRS}[1] {\textcolor{black}{#1}}

\begin{document}

\title{Active colloidal chains with cilia- and flagella-like motion }

\author{S. Gonzalez and R. Soto}

\address{Physics Department, FCFM, Universidad de Chile, Santiago, Chile.}

\begin{abstract}
It has been shown that self-assembled chains of active colloidal particles can present sustained oscillations. These oscillations are possible because of the effective diffusiophoretic forces that mediate the interactions of colloids do not respect the action--reaction principle and hence, a Hopf bifurcation is possible even for overdamped dynamics. Anchoring the particles in one extreme breaks the head-tail symmetry and the oscillation is transformed into a traveling wave pattern and thus the chain behaves like a beating cilium. The net force on the anchor, estimated using the resistive force theory, vanishes before the bifurcation and thereafter grows linearly with the bifurcation parameter. If the mobilities of the particles on one extreme are reduced to mimic an elongated cargo, the traveling wave generates a net velocity on the chain that now behaves like a moving flagellum. The average velocity again grows linearly with the bifurcation parameter. 
Our results demonstrate that simplified systems, consisting only of a few particles with non-reciprocal interaction and head-tail asymmetry show beating motion and self-propulsion.  Both properties are present in many non-equilibrium models thus making our results a general feature of active matter. 
  
\end{abstract}
\submitto{\NJP}

\maketitle

\section{Introduction}

Cilia and flagella are microscopic hair-like structures present in many biological systems. They are used for  propulsion and to stir the surrounding fluid for various vital tasks in the cell~\cite{bray2001cell,fulford1986muco,nonaka2002determination,sawamoto2006new}, all taking place at low Reynolds number. In a nutshell, they are flexible filaments that bend periodically by the action of motors~\cite{gibbons1981cilia,fawcett2014cilia,flagellareview}. For bacteria and other systems, the motors are located on one extreme, in the cell membrane, and the bending of the flagella results from the competition of the viscous and elastic forces. In the case of cilia, the bending is produced by molecular motors situated along all the filament extension, which produce shear stresses that alternate in direction. In both cases, the result is a traveling wave along the filament. This motion is non-reciprocal, meaning that it does not correspond to a rigid body oscillation and hence, it allows to produce propulsion or stirring in the low Reynolds number regime~\cite{purcell1977life}.

Active Janus colloids have emerged as prototypes of non-biological microscopic self-propeling particles. When immersed in a water peroxide solution, catalytic reactions take place at their surfaces, generating motion on the colloid due to phoretic forces~\cite{ActiveColloids}. Recent studies in chains of Janus particles demonstrate the ability of colloidal chains to beat periodically and self propel. This has been observed experimentally in externally~\cite{Nishiguschi2017} and internally driven~\cite{Vutukuri2017} systems. These results suggest a new propulsion mechanism to be synthetically realized: cilia- and flagella-like micro-motors. However, the use of polar particles provides an additional difficulty as they are more complex to manufacture and assembly than isotropic particles.

In this article we take inspiration from self-assembled molecules made of isotropic active colloids~\cite{SG1} to investigate the possibility of beating motion for a system of non-polar particles. We ask the question of whether it is possible that self assembled structures can behave as beating motors and under what constrains this is possible. The answer is affirmative and we explain what is the mechanism behind it: a Hopf transition similar to the one found on active filaments~\cite{Camalet1999}.  However, in contrast to those, there is no time dependent activity nor elastic force between the composing particles. The beating dynamics emerges from static force interactions due to diffusiophoresis and the geometry of our chains, and produce sustained periodic beating for a wide range of parameters in a very simple system. 
By showing what are the minimal requirements to produce sustained beating in an active colloidal system, we hope to give a step in the understanding of synthetic swimmers that mimic yet another feature of biological systems. 

\section{Colloidal model}
\begin{figure}[htp]
\begin{center}
\includegraphics[width=.5\columnwidth]{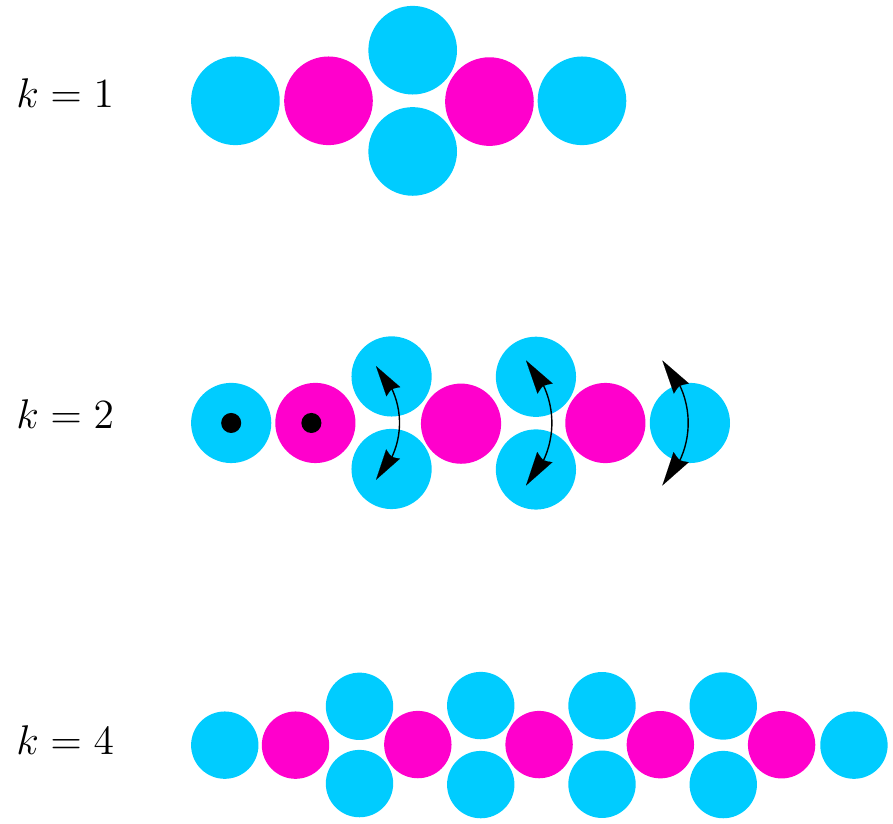}
\end{center}
\caption{Reference chain configuration for different values of $k$. Color indicates  the type of the particles: A in magenta and B in cyan. For $k=2$, the particles with lowered mobility are labeled with  black dots and the arrows indicate the soft degrees of freedom.}
\label{fig.chains}
\end{figure}

For the construction of the cilia and flagella-like structures, we consider a simple model of active particles, which do not present self-propulsion on their own, but rather activity emerges as a result of their interactions. Active colloidal particles, for example gold or platinum particles in a water peroxide solution, generate or consume solutes at their surfaces, which later diffuse in the fluid. For micron sized particles, the contrast of diffusivities between the solute and colloidal particles is large. Hence, the solute concentration profiles are completely enslaved to the instantaneous positions of the colloids $\vec r_i$ and the production rates at their surfaces $\alpha_i$~\cite{RaminModel}. 
In presence of concentration inhomogeneities, a colloid experiences a phoretic drift which is well described by a velocity proportional to the concentration gradient times a mobility coefficient $\mu_i$~\cite{Anderson}. In the far field approximation, the  interaction between two co\-llo\-ids decays as the distance squared and the resulting diffusiophoretic velocity of the colloid $i$ interacting with colloid $k$ is $\vec V_i = U_0\mu_i \alpha_k \hat{r}_{ki}/r_{ki}^2$, where $\vec r_{ki}=r_{ki}\hat{r}_{ki}=\vec r_i-\vec r_k$ is the vector that joins  the centers of the particles and $U_0$ depends on the radius of the particles and the diffusion coefficient of solutes~\cite{SG1}.  Notably, the interaction between two dissimilar particles is not symmetric, violating the action-reaction principle. This is possible by the presence of the supporting fluid that takes away or provides momentum. Besides this interaction, colloidal particles display steric repulsion, which we model as a soft-sphere potential, resulting in the following equations of motion 
\begin{equation}
\frac{d\vec{r}_i}{dt} = \vec V_i, \label{eq.eqmotion0}
\end{equation}
with
\begin{equation}
\vec V_i = U_0 \sum_{k\neq
  i}\alpha_k\mu_i\frac{\hat{r}_{ki}}{{r}_{ki}^2} +  U_1\sum_{k\neq
  i}\frac{\hat{r}_{ki}}{{r}_{ki}^{12}} , \label{eq.eqmotion1}
\end{equation}
where $U_1$ measures the intensity of the repulsive potential. The combination of these two interactions allow that stable clusters self-assemble. For simplicity, no Brownian noise is added.

In Ref.~\cite{SG1} the case of a binary mixture (of types $A$ and $B$) was studied in detail. Characterizing the diffusiophoretic charges $\alpha$ and $\mu$ by their signs, it was shown that a particularly interesting case occurs when $\alpha_A,\mu_A>0$ and $\alpha_B,\mu_B<0$. Consequently, equal particles repel and dissimilar particles attract with an intensity that depends on the values of their charges. Under appropriate conditions, this attraction allows the formation of clusters where $A$ and $B$ particles alternate forming chains. As neighbor particles are different, the effect of the violation of action-reaction is  stronger and it is more possible that activity emerges. In the  cases when the chain presents a head-tail asymmetry (for example, in the trivial $AB$ cluster), activity  takes the form of self-propulsion. Normally, unless the chain is perfectly symmetric, a residual propulsion velocity will result, given by the combination of the geometry and the values of the charges. Also, if the chain presents some chiral asymmetry it will spontaneously rotate~\cite{SG1}. 

\comRS{Although the equations are derived, and valid, for a three dimensional system, we restrict the movement to two dimensions for two reasons: first, most of the experimental realizations of colloidal systems are not density matched, thus the particles tend to sediment and their final motion occurs in a plane~\cite{Simmchenn2016,Niu2017}. Secondly, the analysis of the geometrical structures that are formed is simpler in two dimensions, without unnecessary complications from the increased number of degrees of freedom in three dimensions. Nevertheless, the principles and qualitative results obtained here are applicable also in three dimensions.}

\begin{figure}
\centering
\includegraphics[width=.49\columnwidth]{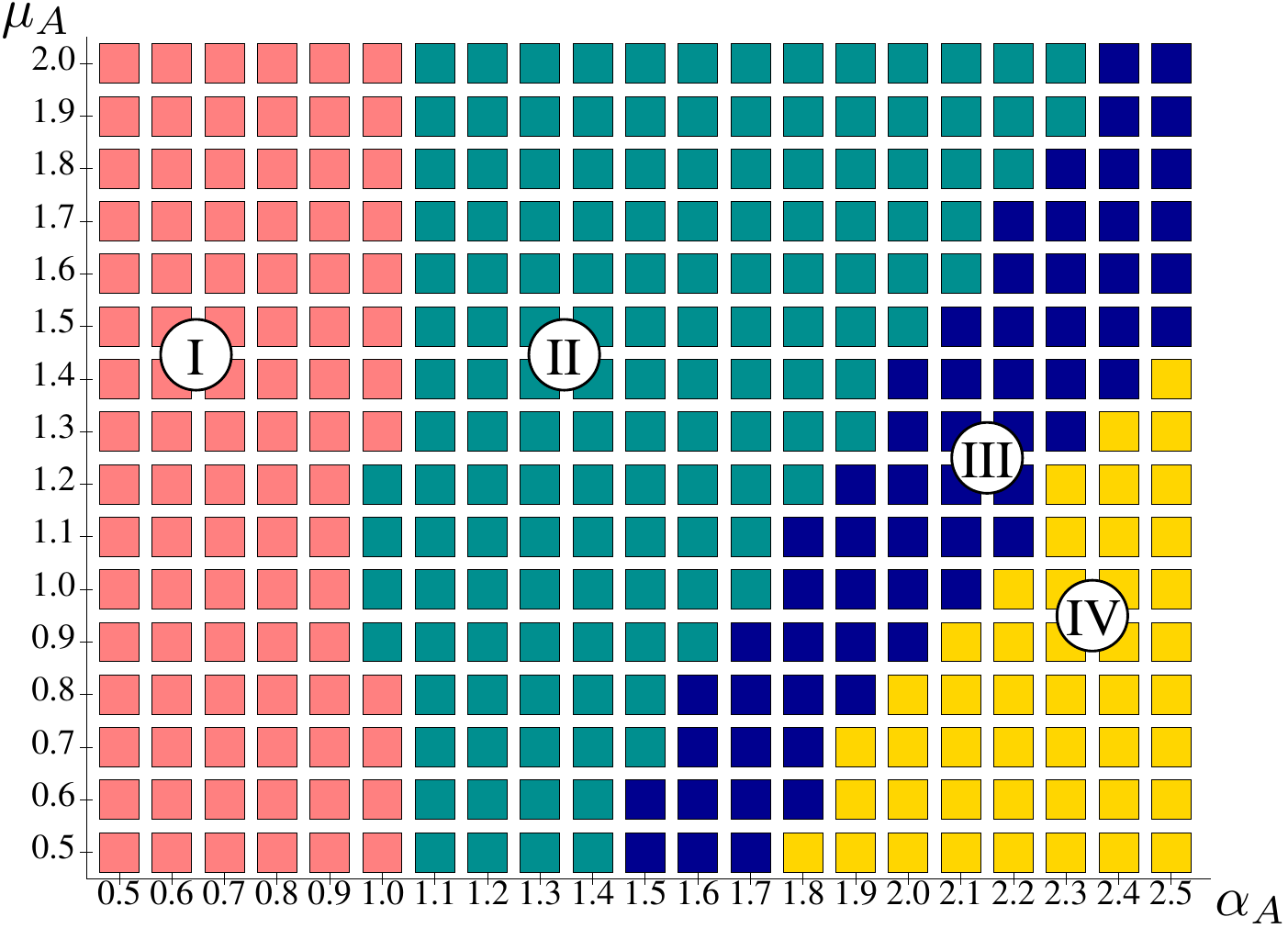}
\includegraphics[width=.49\columnwidth]{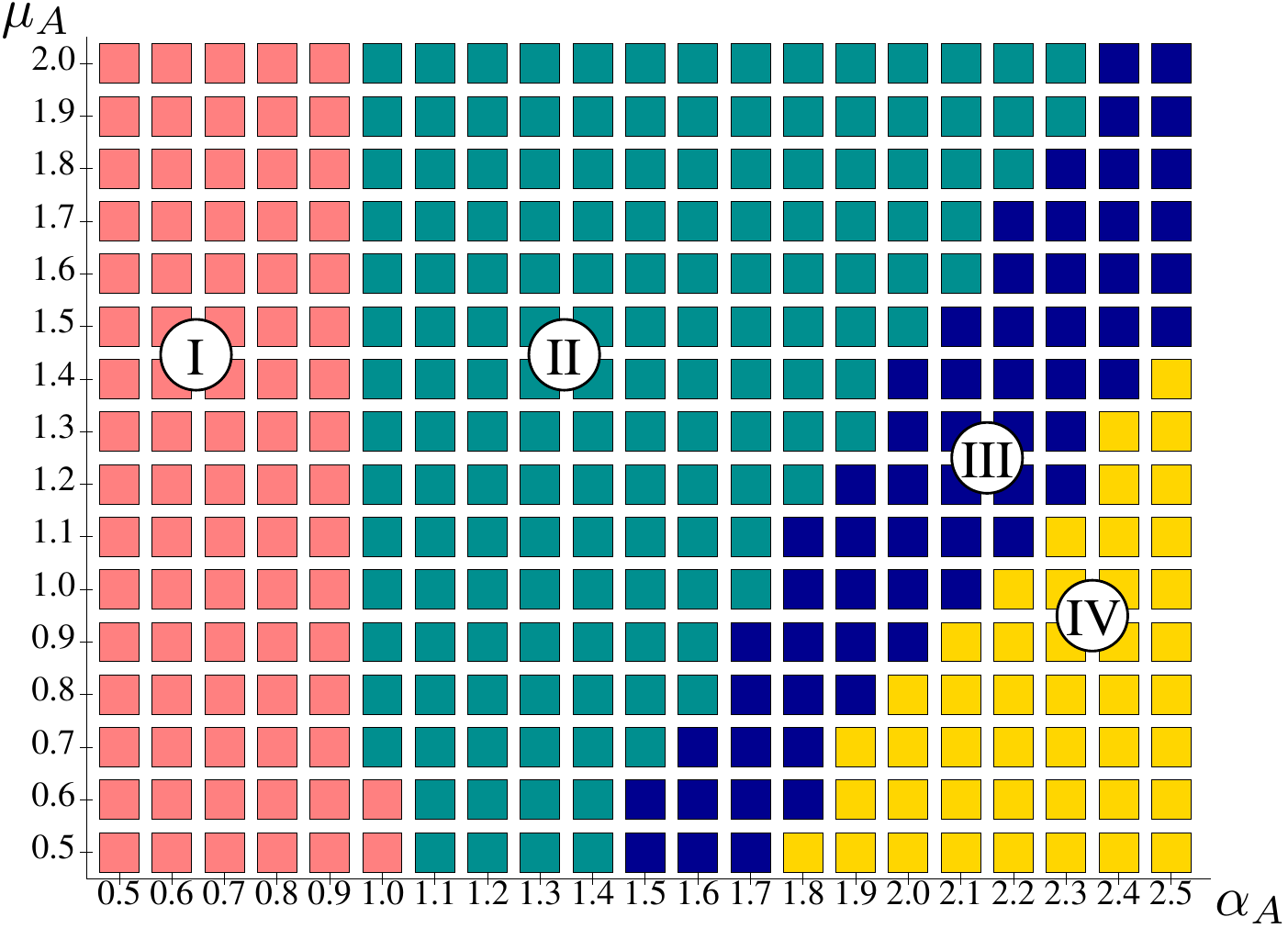}
\caption{Phase diagram for anchored chains (left) and chains with a cargo (right) in the $\alpha_A$-$\mu_A$ space for $k=2$. Four phases are identified: chains can be unstable by disintegration (I, pink), linear chains are stable (II, green), linear chains oscillate periodically (III, blue), and chains are unstable by collapse (IV, yellow).}
\label{fig.phasediag}
\end{figure}

In a chain another kind of activity can appear besides self-propulsion and rotation, and is the subject of this article. Despite the fact that the particle dynamics is overdamped, that is without inertia (Eq.~\ref{eq.eqmotion1}), as the interactions are not symmetric the global dynamics is not necessarily variational. When the linear stability of a colloidal molecule is studied, the violation of the action-reaction principle produces an interaction matrix that is not symmetric and complex conjugate eigenvalues can appear. Thus, it is possible that by moving the interaction parameters the molecule becomes unstable via a Hopf bifurcation~\cite{Hopf}. The system then evolves toward a limit cycle, oscillating periodically. In Ref.~\cite{SG2}, the case of a symmetric oscillator was studied, showing that there is even a route for self-assembly and that the oscillation is robust under thermal noise. 

Here, we consider linear chains as those presented in Fig.~\ref{fig.chains}, which are labeled by the number of units $k$, being composed of $3(k+1)$ particles. These chains are stable for a wide range of parameters with the combinations of charge signs described above. It is possible to absorb the dimensions of $\alpha$ and $\mu$ into $U_0$, allowing us to chose  $\alpha_B=-1$ and $\mu_B=-1$, leaving $\alpha_A$ and $\mu_A$ as free variables. Also, time and space units can be chosen such that $U_0=U_1=1$, implying that the minimal distance between particles that attract is order one, fixing therefore the hard-core diameter.

Due to their symmetry, these chains cannot self-propel. Also, when the Hopf bifurcation is crossed, the limit cycle corresponds to a global phase and the chain  oscillates in a normal-mode form (in Ref.~\cite{SG2} the case with $k=3$ is studied). Here, we aim to introduce a simple asymmetry which can generate traveling waves along the chain, and, eventually the possibility of motion or thrust on a cargo, similarly to the oscillation of flagella and cilia. 
To create a head-tail asymmetry in the chain, we keep the geometry and distribution of phoretic charges and change only the mobility of the colloids on one extreme (see Fig.~\ref{fig.chains}). 

\section{Anchored chains: cilium-like motion}
\begin{figure}
\centering
\includegraphics[width=.7\columnwidth]{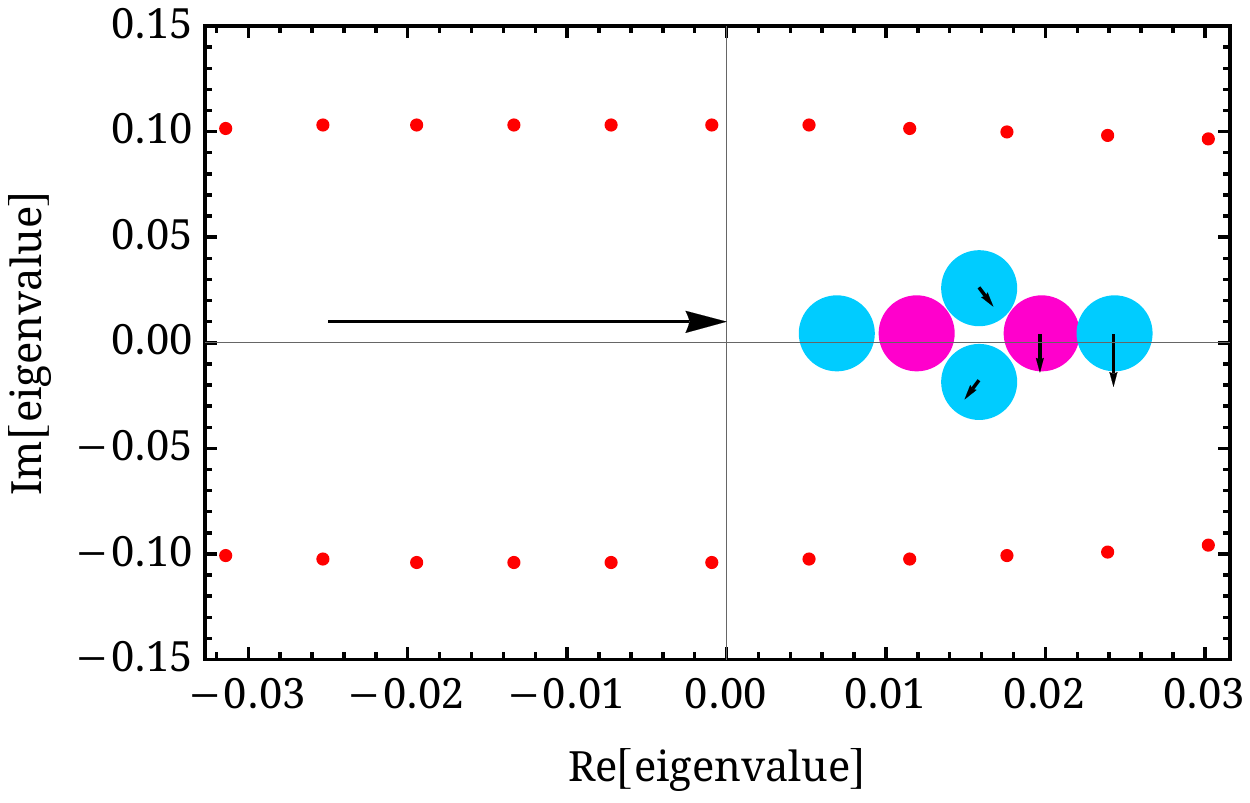}
\begin{picture}(10,10)(0,0)
\put(-190,130){$\alpha_A$}
\end{picture}
\caption{\comRS{Real and imaginary parts of eigenvalues associated to the soft modes for an anchored chain with $k=1$, when $\alpha_A$ is varied in the range $[1.62,1.72]$ (indicated by an arrow) and $\mu_A=1$ is fixed. The critical value for the transition, when the real parts become positive, is $\alpha_{Ac} \simeq 1.67$. At the inset, the associated eigenmode is depicted with arrows. There are other six eigenvalues, which are purely negative and real, and of much higher magnitudes, representing the relaxation of the hard degrees of freedom of the chain. To compute the eigenvalues, for each $\alpha_A$ the equilibrium (eventually unstable) configuration is found by relaxing for a finite time a perfectly symmetric chain. Then, the force matrix is computed by performing linear perturbations to all degrees of freedom.}
}

\label{fig.eigenvalues}
\end{figure}
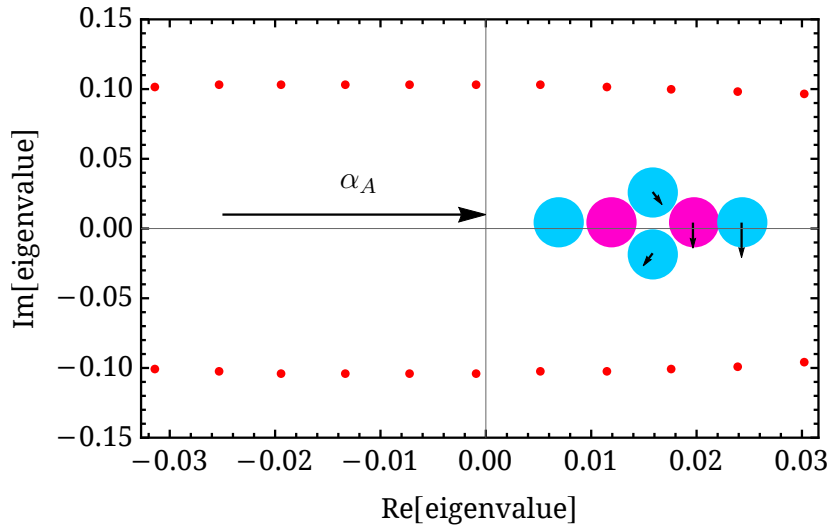
As a first case, we reduce the mobility of the two leftmost colloids to zero, modeling a chain that is rigidly anchored (for example using optical traps). \comRS{Alternatively, this could be realized, for example, by  using a fixed Janus particle as the head of the structure.} Note that it is not enough to anchor one colloid as the chain would have kept a global rotational degree of freedom that obscures the analysis. Although in a chain there are many degrees of freedom, most of them are frozen by the balance of the  attraction by the phoretic forces and the hard core repulsion. These hard degrees of freedom are almost fixed to the equilibrium distance and can be treated as constrains~\cite{SG2}. The rest are soft modes, which for the chains correspond to rotations as presented in Fig.~\ref{fig.chains}. The number of soft degrees of freedom is $g=k+1$.

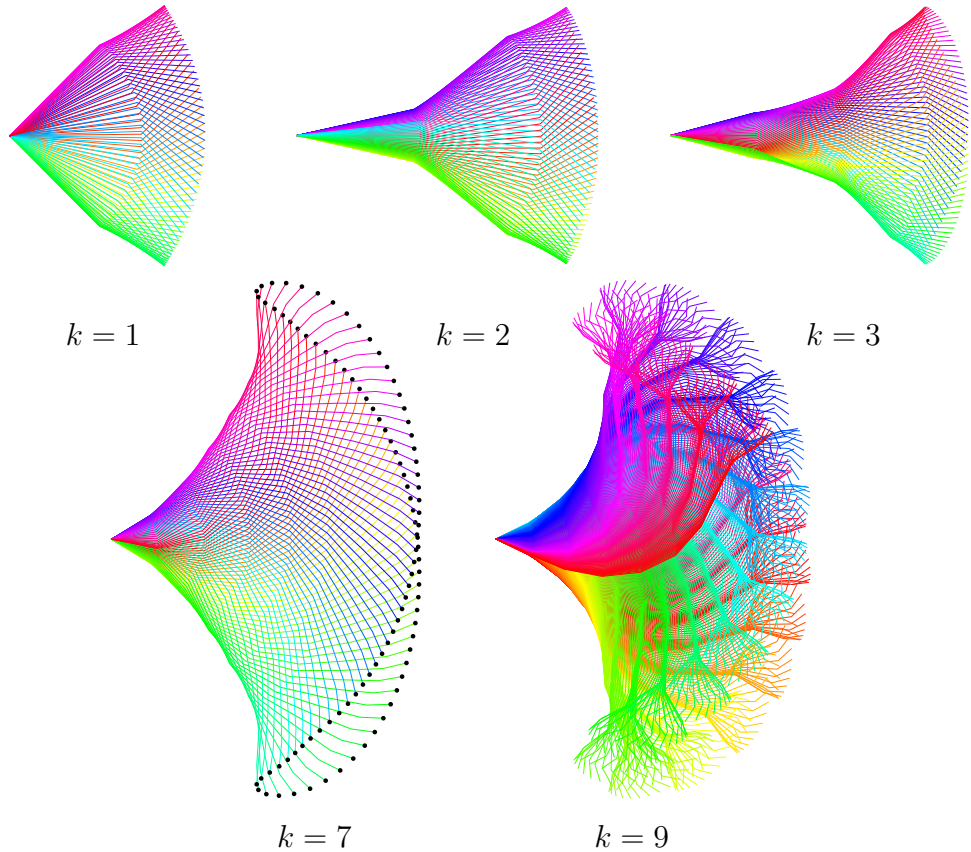
\begin{figure}
\centering
\includegraphics[height=.23\columnwidth]{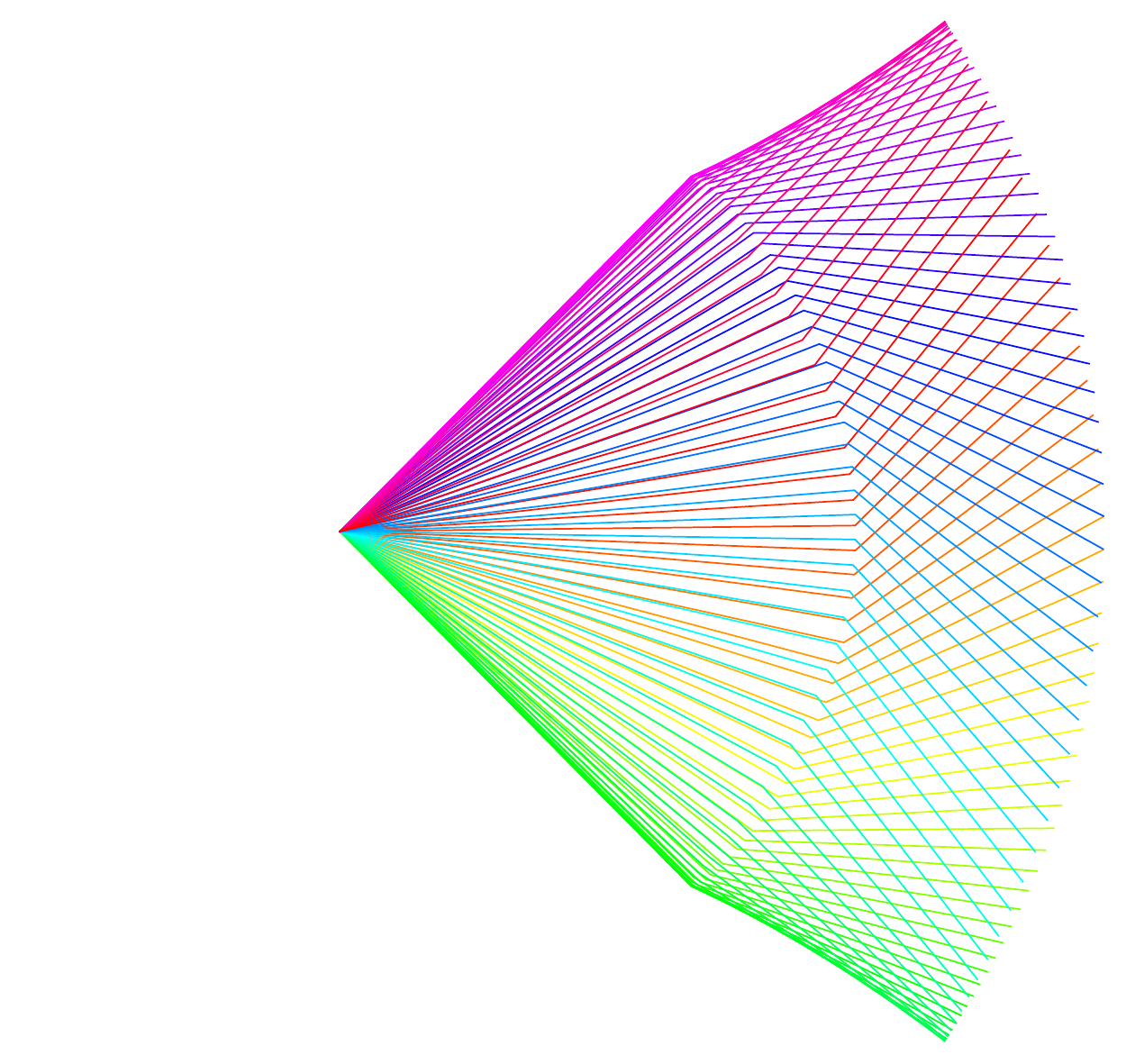}
\includegraphics[height=.23\columnwidth]{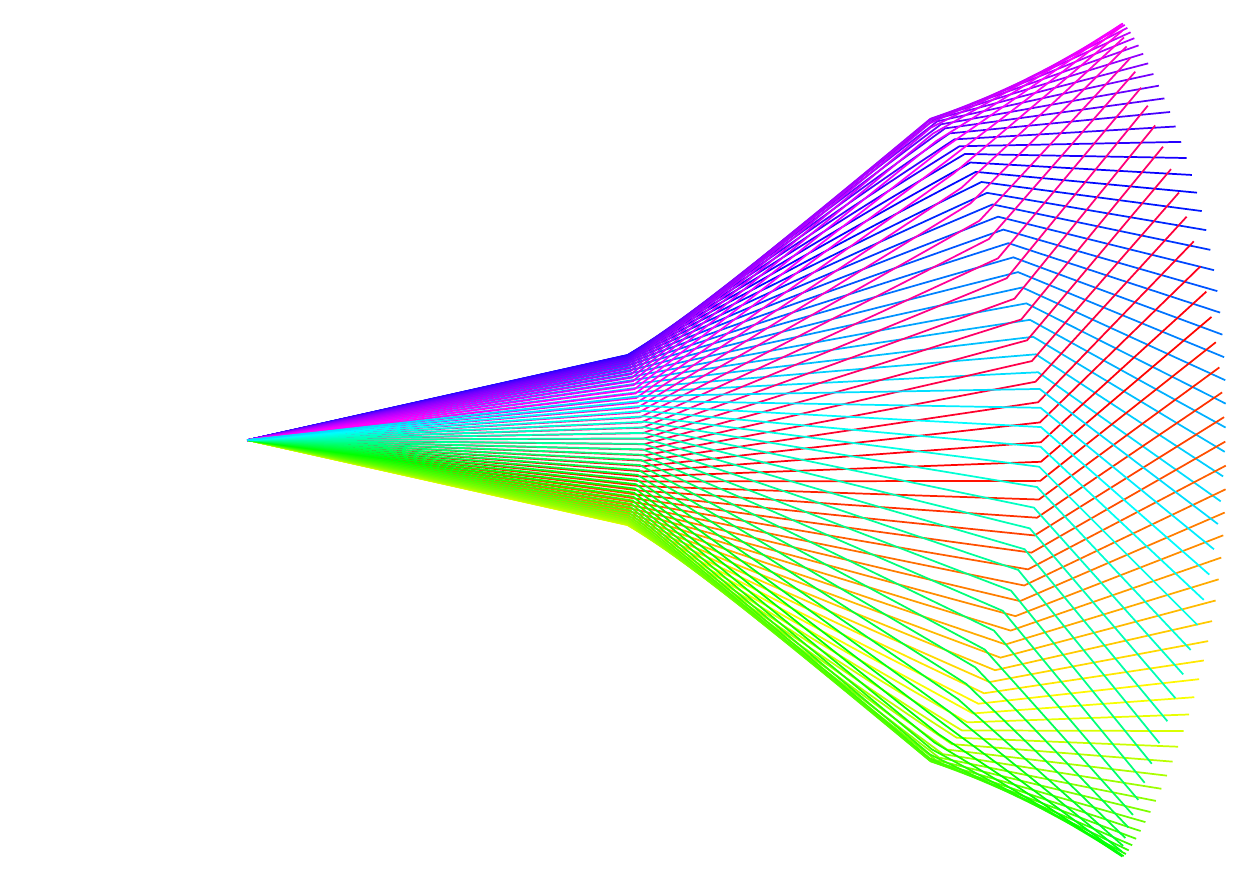}
\includegraphics[height=.23\columnwidth]{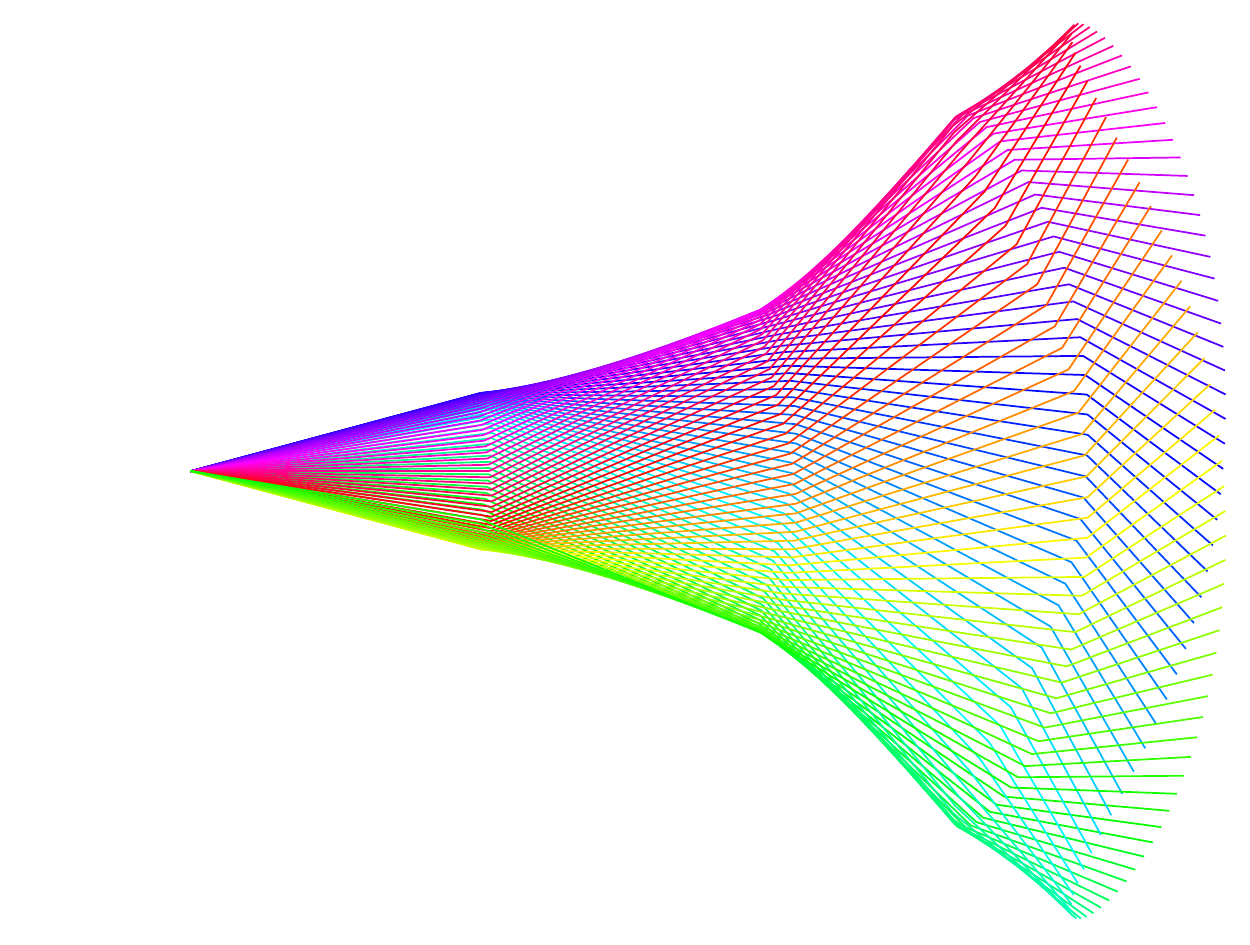}

\includegraphics[height=.45\columnwidth]{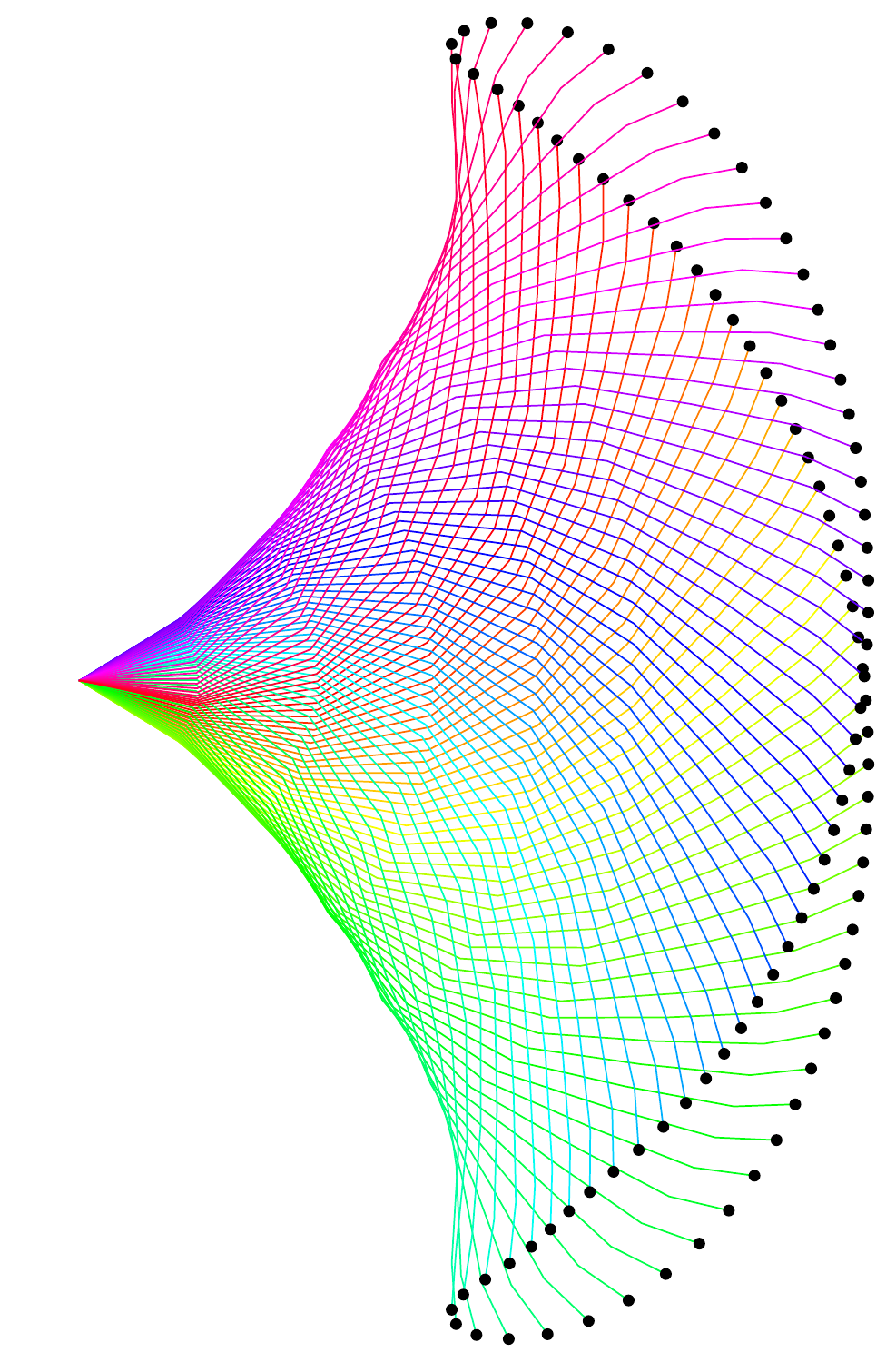}
\quad
\reflectbox{\rotatebox[origin=c]{180}{\includegraphics[height=.45\columnwidth]{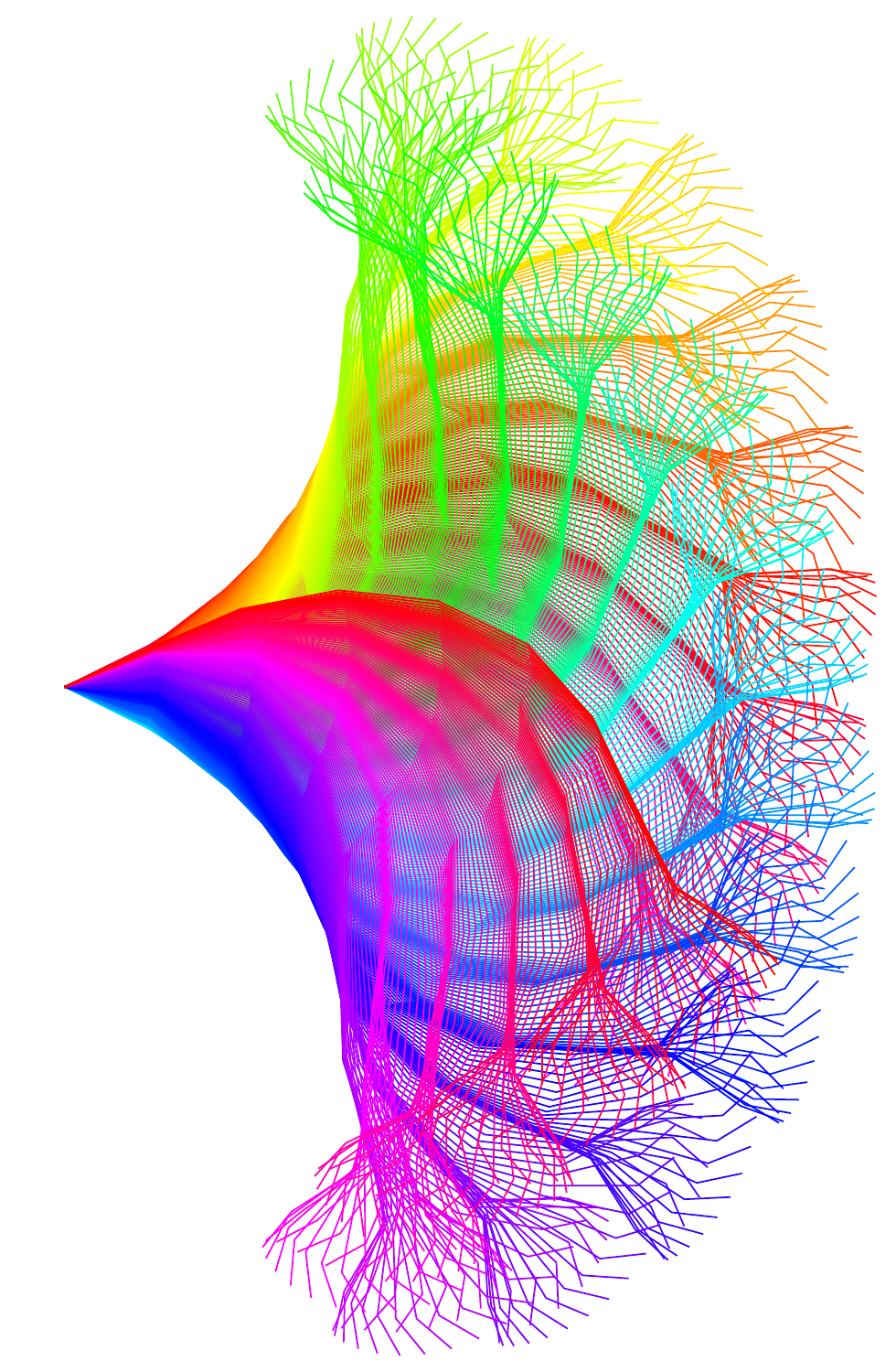}}}

\begin{picture}(10,10)(0,0)
\put(-140,190){$k=1$}
\put(-0,190){$k=2$}
\put(140,190){$k=3$}
\put(-60,0){$k=7$}
\put(60,0){$k=9$}
\end{picture}
\caption{
Geometry of the flagellar beating patterns of anchored chains for different values of $k$, with  lines joining the backbone particles. Colors indicate  different time instants covering one period, with time increasing in the sequence: green, cyan, blue, magenta, red, yellow, and back to green again. The parameters in all cases are $\mu_A=0.5$ and $\alpha_A=1.5$ except for $k=9$, where  $\mu_A=0.25$.  For all chains, by decreasing $\mu_A$ the amplitude and frequency of oscillation grow and in the case of the $k=9$ chain a secondary oscillation is produced, possibly due to a secondary Hopf bifurcation. For $k=7$, black dots on the last particle help to visualize the non-reciprocal motion. For $k=9$, the sampling rate has been increased to highlight the fast oscillation of the rightmost particles.  (See Supplementary Video 1.)}
\label{fig.beating}
\end{figure}

To obtain stable oscillations via a Hopf bifurcation, a pair of complex conjugate eigenvalues with positive real part must exist when the linear stability analysis of the chain is performed. For this, it is necessary to have at least two soft degrees of freedom, which in our case is obtained even for $k=1$, as it is indeed the case, see Fig.~\ref{fig.beating}.
Figure~\ref{fig.phasediag} presents the phase diagram, in the $\alpha_A$-$\mu_A$ space for $k=2$, obtained by analyzing the final state of chains prepared initially in the linear configuration plus a small perturbation. Various states are obtained: for small values of $\alpha_A$ (region I) the chains are unstable by disintegration due to the large phoretic repulsion between equal particles, in region II the chains are stable in their linear configuration, and in region III stable oscillations take place. The amplitude of the oscillations grows when moving from region II to III, as the square root of the bifurcation parameter (see below). When this amplitude becomes large enough, the chain can fold into itself and collapses into a compact cluster, which no longer oscillates (region IV). Analogous behaviors are obtained for other values of $k$. 

In order to analyze in more detail the properties of the oscillators, we extend their stability by adding springs between colloids that are initially neighbors (i.e., in the linear configurations displayed in Fig.~\ref{fig.chains}). In Eq.~(\ref{eq.eqmotion1}) appropriate terms of the form $\hat{r}_{ki}(r_0-|\vec{r}_{ki}|)$ are added, where $r_0$ are the equilibrium distances, reinforcing the hard degrees of freedom but not altering the soft ones. In this way, under large oscillations, the phoretic forces are not able to break or fold the chain, which keeps its structure. With this method, region IV disappears and stable oscillations are obtained instead. At this point, it is worth mentioning that although here the analysis is done for the case of diffusiophoretic forces, the principles described in this article are valid for any system presenting effective interactions that do not satisfy the action reaction principle. This is the case, for example, of interactions mediated by wakes, shadows or complex solvents, or even for Casimir-like forces (see Ref.~\cite{NonActionReaction} and references therein). Under these conditions activity emerges naturally as a result of interactions. In this context, the addition of the cohesive spring forces must be understood as any effective or direct interaction that can keep the structures stable. Experimentally, at least one method exists to create elastic bonds between the Janus particles: using electric fields to stabilize the geometry, a heating step to bond the chains and finally sterical stabilization to make them flexible \cite{Vutukuri2017,Vutukuri2012}.

\begin{figure}
\centering
\includegraphics[width=.7\columnwidth]{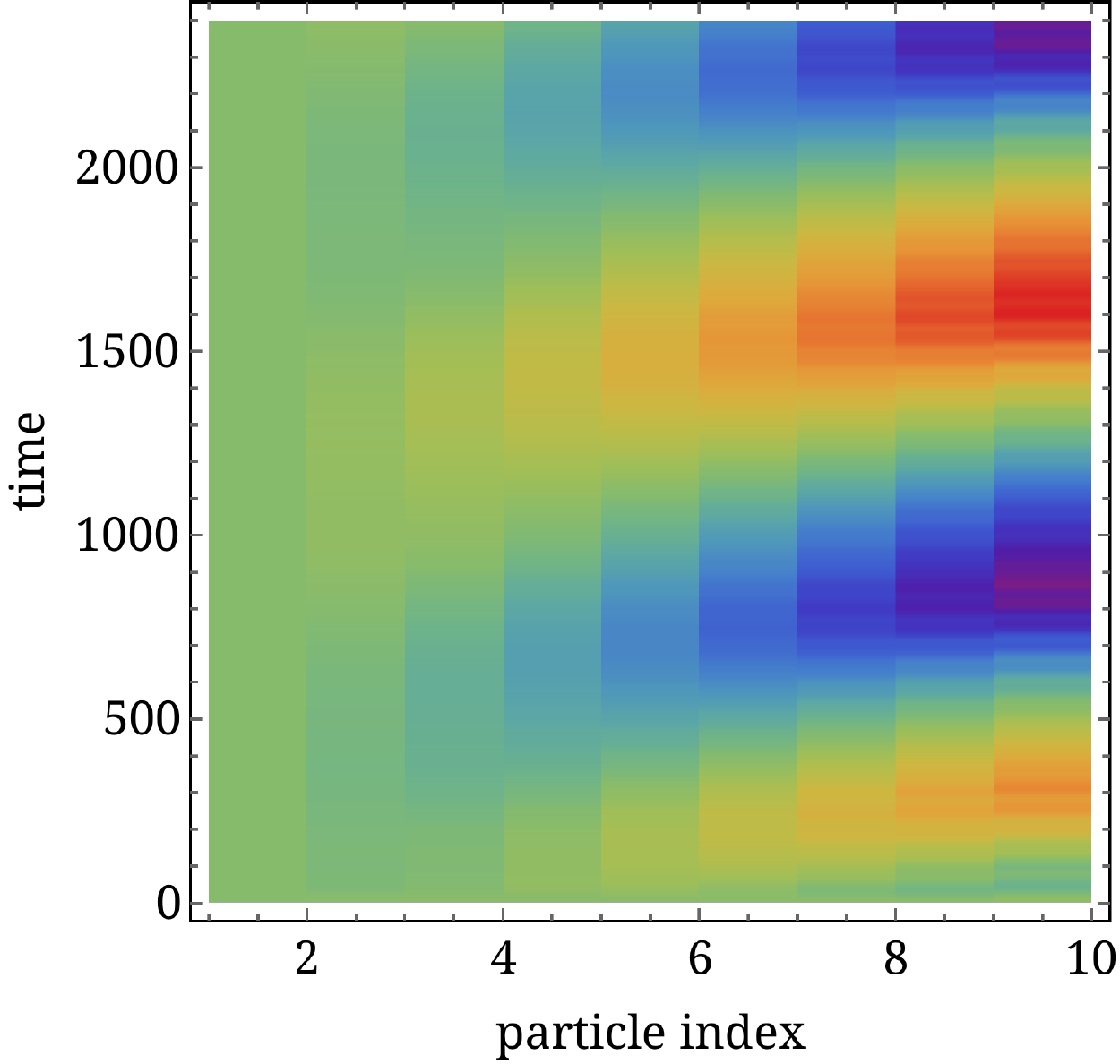}
\caption{Spatio-temporal diagram of the vertical displacement of the backbone particles for an anchored chain of  $k = 9$, $\alpha_A=1.5$, and $\mu_A=0.25$. Note how higher harmonics are evident in the motion of particles close to the free end (high particle index).
}
\label{fig.travelwave}
\end{figure}

Figure~\ref{fig.beating} presents the beating patterns for anchored chains with $k=1$ to $k=9$, where springs have been used for $k=7$ and $k=9$ to keep stable the structures for such high amplitudes. An important feature is that the chains do not oscillate as rigid bodies neither as a normal mode, and the back and forth patterns are not reciprocal. Instead, due to the head-tail asymmetry, a traveling wave appears. Supplementary Video 1 shows the movement of these chains. For large $k$, the existence of this wave is more evident in a spatio-temporal diagram of the vertical displacements, as shown in Fig.~\ref{fig.travelwave}, where it is clear that a wave propagates outward. Essentially, the wavelength is fixed by the chain length, except that for $k\geq9$  it is possible to see that the rightmost particles oscillate at a higher frequency, reflecting the existence of higher harmonics in the limit cycle. For small chains, the existence of the traveling wave is more clear when the evolution of the soft degrees of freedom is analyzed. For example, for $k=1$, the two bending angles converge to a limit cycle, with a phase shift that in this case represents also a wave that travels in the outward direction (not shown).

\begin{figure}
\centering
\includegraphics[width=.425\columnwidth]{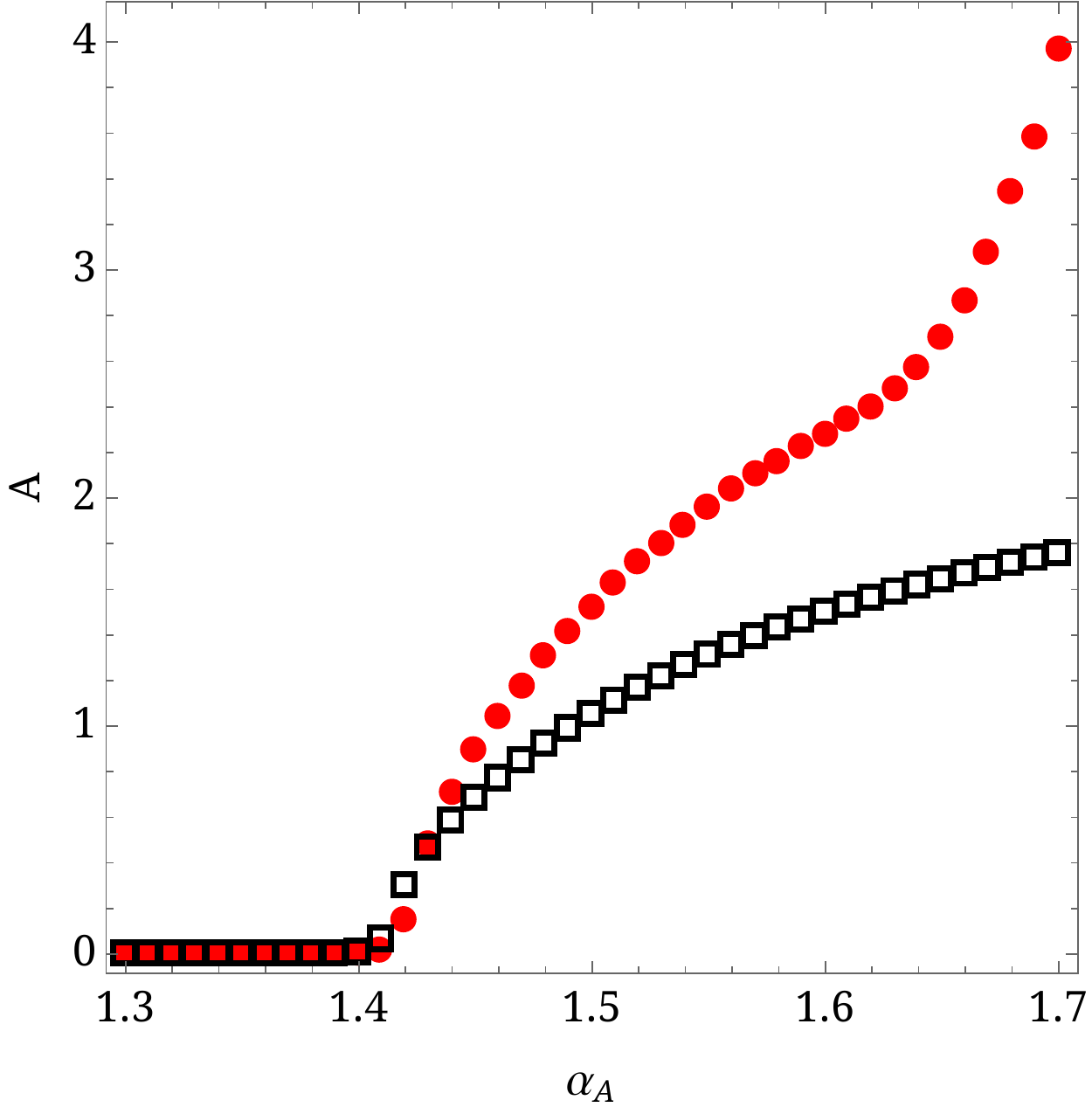}
\includegraphics[width=.45\columnwidth]{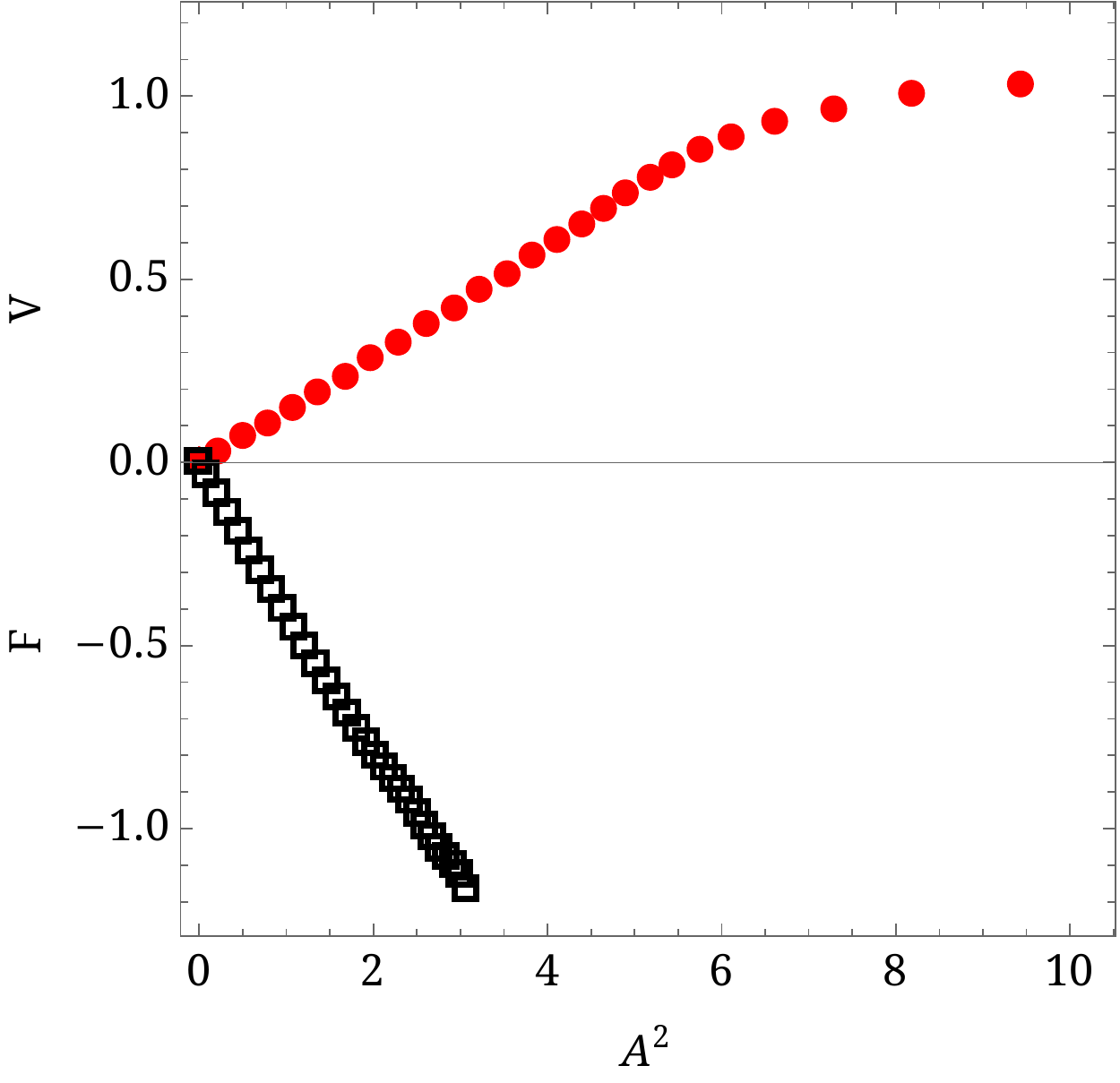}
\caption{
Left: Oscillation amplitude as a function of $\alpha_A$ for an anchored chain (open black squares) and for a chain with a cargo (solid red circles). The critical charge for the anchored chain is $\alpha^{\rm c}_A=1.406$  and for the chain with a cargo is $\alpha^{\rm c}_A=1.413$. The amplitude is computed for the vertical component of the center of mass as the difference between the highest and lowest point of its trajectory after have reached the steady state. 
Right: Average force on the head  for an anchored chain (open black squares) and average velocity for a chain with a cargo (solid red circles) as a function of the respective amplitudes squared. In all cases, $k=2$ and $\mu_2= 0.5$. Note that the amplitude grows as the square root of the transition parameter $\alpha_A-\alpha_A^{\rm c}$, until the secondary oscillation appears around $\alpha_A =1.65$; cf.\ Fig.\ \ref{fig.secondary}}
\label{fig.amplitude}
\end{figure}

The existence of a traveling wave implies that the motion is not reciprocal and propulsion can take place (see Ref.~\cite{purcell1977life} for a discussion on the role of nonreciprocal motions to generate thrust and net displacement in the low Reynolds regime). As the head is anchored, it is not possible for the chain to self-propel but rather it can exert a net force on the head. To estimate this force, we use the resistive force theory~\cite{Flagella1976,brennen1977fluid,johnson1979flagellar}. For this, a spline curve of degree 2 is generated at each instant using the centers of particles $A$. This curve is then divided in 100 intervals, and on each, the perpendicular and tangential vectors to the curve are computed, allowing to project the velocity of each interval into this base. The force in the  infinitely thin  approximation is computed as $\vec F = R_0 \sum_n (2 \vec v^{\rm perp}_n+ \vec v^{\rm tang}_n)$, where the sum runs over all the intervals and $R_0$ is the unitary resistance, which is proportional to the fluid viscosity. Finally, this force is averaged over a full cycle. 
Figure~\ref{fig.amplitude} presents the net force on the head as a function of $\alpha_A$. The force is proportional to the oscillation amplitude squared and takes negative values, that is pointing towards the head. This result is consistent with analytic calculations for the force generated by a sinusoidal wave traveling outward on a flagellum, which also scales with the amplitude squared and points to the head~\cite{Flagella1976}. An important feature, which is part of the design of the chain, is that the force vanishes by symmetry when the chain is in equilibrium and it pushes only as a result of the oscillations. Notably, the oscillations are not driven by an external motor, but rather they are spontaneously generated by the chain when parameters are moved.

The oscillation period is almost independent of the control parameters $\alpha_A$ and $\mu_A$. This is consistent with the fact that the oscillation results from a Hopf bifurcation, where the real part of a pair of eigenvalues change sign, but the imaginary part is a smooth function of the parameters, with no qualitative change at the transition~\cite{Hopf}.  \comRS{The evolution of the eigenvalues for $k=1$ is shown in Fig.\ \ref{fig.eigenvalues} and has beed previously reported for $k=3$ in Fig.\ 2 from Ref.\ \cite{SG1}.}

\begin{figure}
\centering
\includegraphics[width=.75\columnwidth]{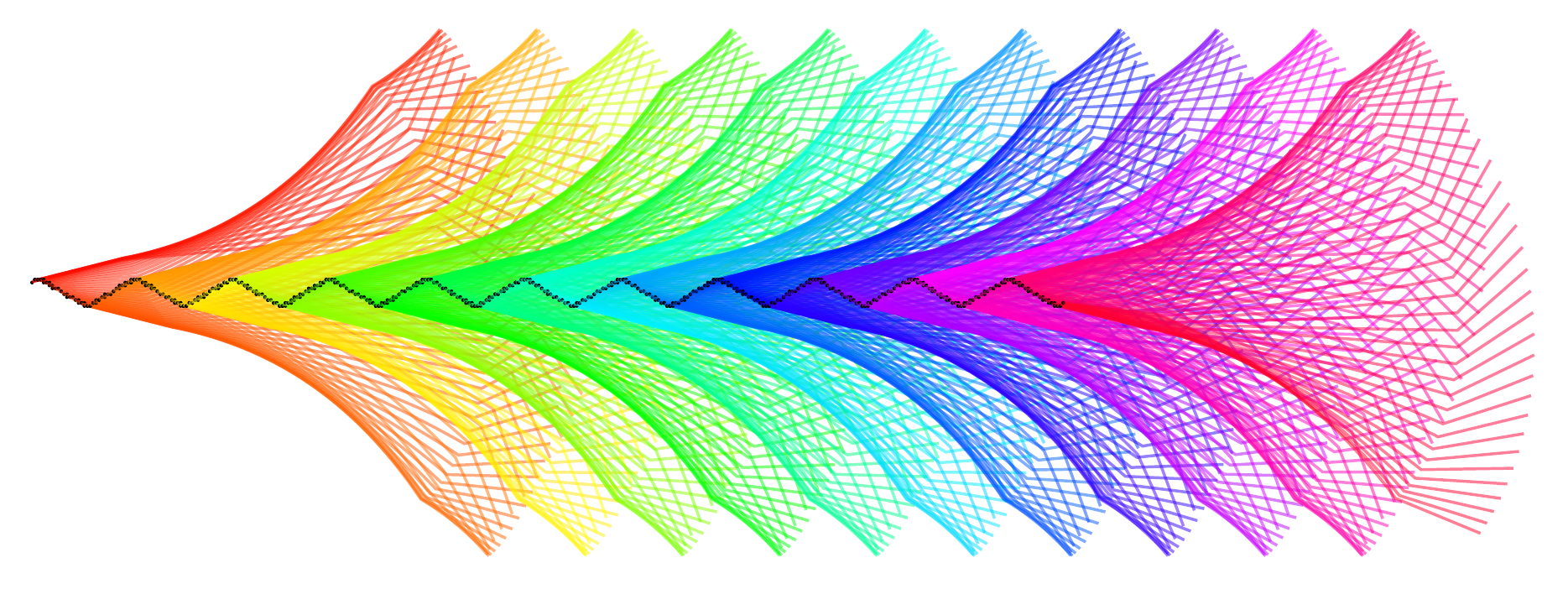}
\includegraphics[width=.75\columnwidth]{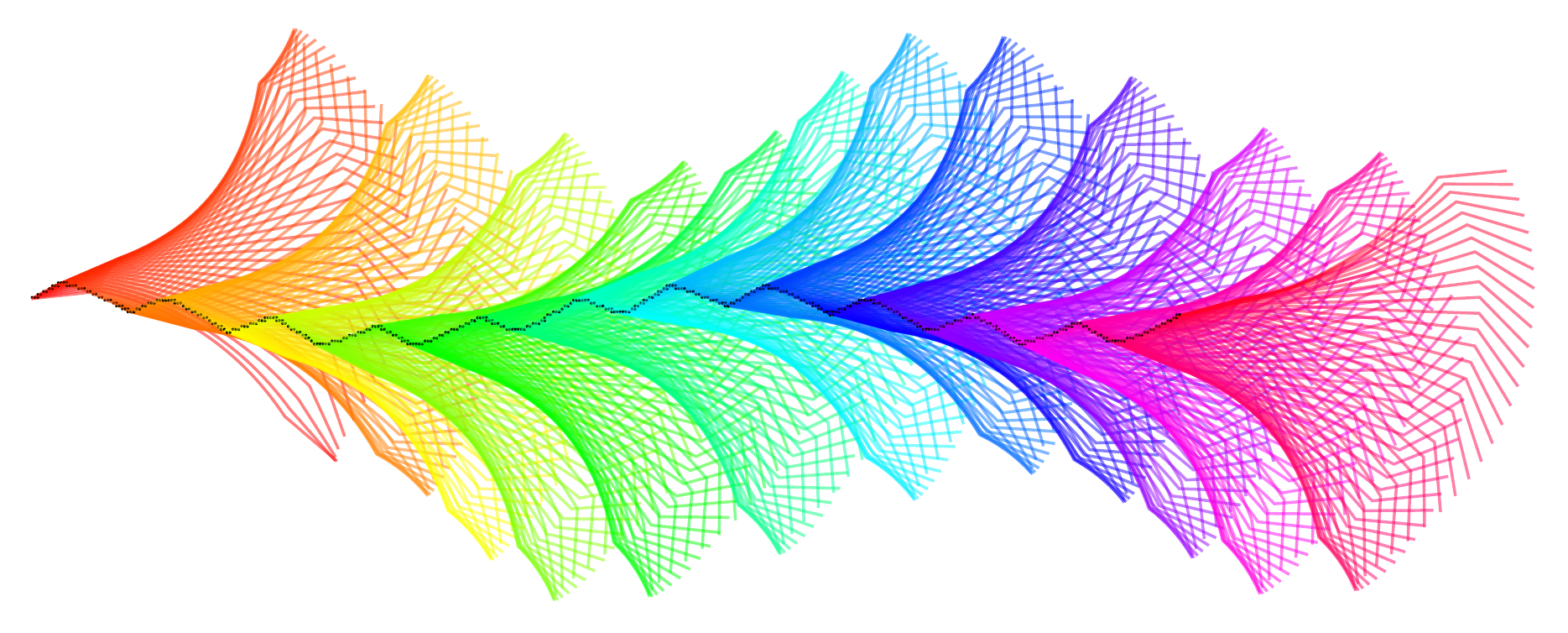}
\begin{picture}(10,10)(0,0)
\put(20,200){$\alpha_A=1.6$}
\put(20,60){$\alpha_A=1.7$}
\end{picture}
\caption{Geometry of the flagellar beating patterns of a chain with a cargo for  $\alpha_A=1.6$ (top) and $\alpha_B=1.7$ (bottom), where a secondary instability takes place in the form of a low-frequency oscillation.
Lines join the backbone particles, with colors indicating the different time instants covering one period of the low frequency oscillation, with time increasing in the sequence: red, yellow green, cyan, blue, magenta, and back to red again. In black, the position of the leftmost particle (flagellum's head).  For $\alpha_A=1.7$, the oscillation frequency is slightly higher, resulting in a larger displacement for the same time. The secondary oscillation for this case is fully developed.}
\label{fig.secondary}
\end{figure}

\section{Chains with a cargo: flagellum-like motion}

\begin{figure}[ht]
\centering
\includegraphics[width=.75\columnwidth]{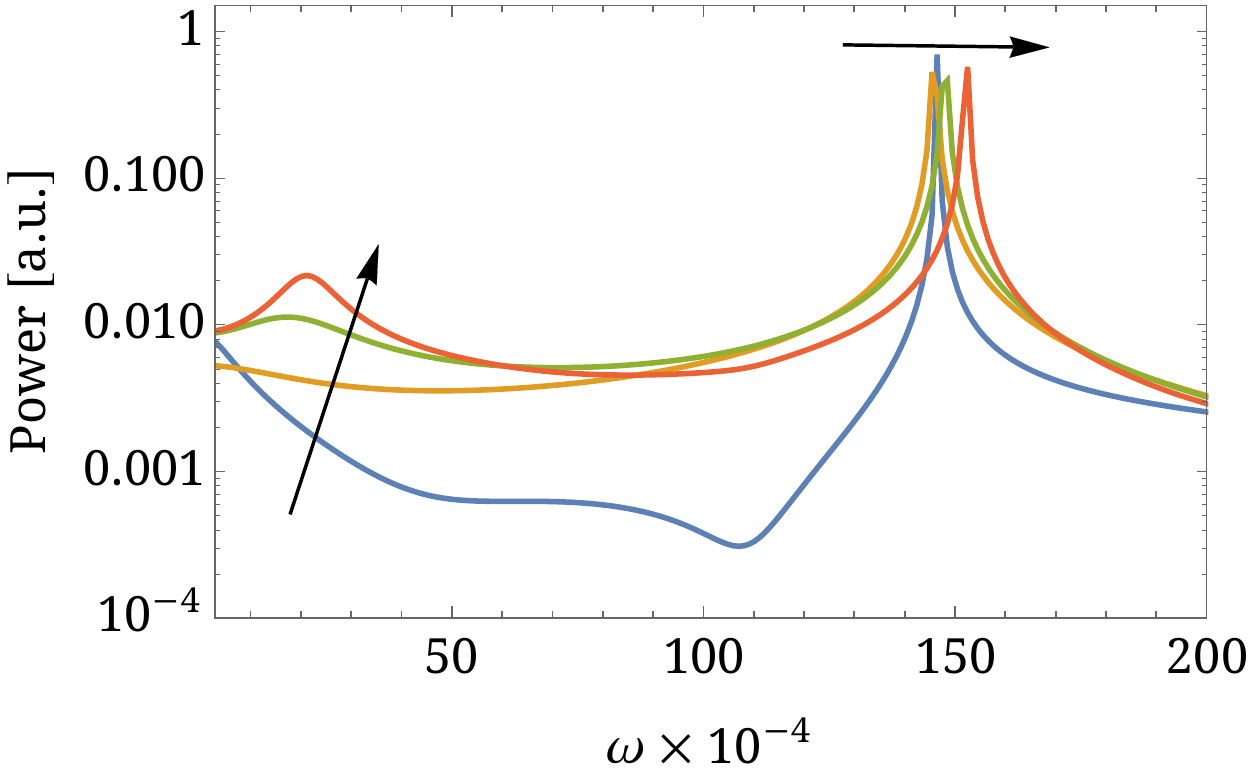}
\begin{picture}(10,10)(0,0)
\put(-250,95){$\alpha_A$}
\put(-125,190){$\alpha_A$}
\end{picture}

\caption{Power spectrum of the vertical center of mass coordinate for $\alpha_A=1.50$, 1.55, 1.60, and 1.65, increasing as indicated by the arrows. The simulations last 10000 time units and are sampled every time unit.
The primary oscillation (largest peak on the right) shifts towards higher values, i.e.\ shorter periods, as $\alpha_A$ increases, and another slow frequency peak at $\omega \sim 30$ appears for $\alpha_A\gtrsim1.6$, with an amplitude that increases with $\alpha_A$. }
\label{fig.power}
\end{figure}

We now aim to  generate a chain that can self-propel when oscillations start but otherwise would remain at rest when it is at the equilibrium shape. For this, we lower the mobility of the two leftmost particles in order to mimic a cargo, like the head of a flagellated microorganism~\cite{flagellareview}.  \comRS{Again, as in the case of the cilia-like movement we need that the first two particles have reduced mobility to avoid the appearance of a global rotational mode.} For an elongated cargo aligned with the principal axis of the chain, the longitudinal and transversal mobilities satisfy $M_l=2M_t$ and we are free to chose $M_l$, which we take equal to 0.2 representing a large head. Then, the equations of motion~(\ref{eq.eqmotion0}) for the first two particles are modified to
\begin{equation}
\frac{d\vec{r}_i}{dt} = \left[M_l \hat n\hat n + M_t (I-\hat n\hat n)\right] \cdot \vec V_i,
\end{equation}
where $\hat{n}=\hat r_{12}$, while we keep the form of $\vec V_i$ in Eq.~(\ref{eq.eqmotion1}). Videos of the dynamics due to different mobilities are found in Supplementary Videos 2, 3 and 4.

Releasing the first two colloids, that now can move, adds three new degrees of freedom. However, these are not relevant for producing additional bifurcations or complexifying the attractors in phase space, because they are associated to a global rotation and translation of the chain. In summary, for these chains with cargo, the number of soft degrees of freedom is again $g=k+1$ and oscillations can emerge via Hopf bifurcations at $k=1$. Simulations of the system show a phase digram very similar to that of the anchored chain, with the difference that oscillations are accompanied with a net translation (see Fig.~\ref{fig.phasediag}). Notably, the average velocity due to the phoretic drift is positive and the chain moves towards the tail with a velocity that grows as the amplitude squared. Again, by symmetry, if the chain is not oscillating, there is no net velocity. If hydrodynamic interactions are included and modeled with the resistive force theory, a net force toward the head is generated as shown in the previous section which would give rise to a velocity in that direction. Both velocities are proportional to the amplitude squared but point in opposite directions.  The net velocity on an experimental condition would depend on the balance of the phoretic charges, the viscosity and the particle sizes. As this is a proof of concept proposal, where many details which depend on the particular realization for the experiment can not be foreseen, it is futile to predict the direction of motion. Note, however, that as both velocities have the same dependence with the amplitude, it is not expected that a crossover takes over in the vicinity of the bifurcation line. 

Figure \ref{fig.amplitude} shows the velocity of displacement for a cargo made with $k=2$ and $\mu_A=0.5$ and varying $\alpha_A$. The Hopf bifurcation for the flagellum and cilium cases take place at very similar values of $\alpha_{A}\approx 1.41$. For both cases, the amplitude of oscillation increases as the square root of the bifurcation parameter, however for the flagellum case, a secondary oscillation appears at $\alpha_A\gtrsim1.6$ and makes the amplitude of oscillation larger (see Fig.~\ref{fig.secondary}). It must be noted that the frequency of oscillation is almost constant while varying $\alpha$ (see Fig.~\ref{fig.power}) so the increase in speed is only due to the shape of the beating pattern that changes with $\alpha_A$. Figure \ref{fig.power} shows that, just as for the anchored chain, the cargo's  main beating frequency (largest peak) only slightly increases as increasing $\alpha_A$. The secondary slow oscillation, in contrast, is inexistent for $\alpha_A=1.55$ and increases for larger $\alpha_A$.

\section{Discussion}

We have shown a series of chain-like structures composed of spherical active colloids that can behave either as a cilia or a flagella. These structures can in principle self-assemble, and in fact the ``reaction path'' for at least one of them is known \cite{SG2}. The (self-assemblable) chains are stable in a wide range of the parameter space and for which the building blocks (dimers and trimers) have been already experimentally realized \cite{Niu2017}. These building blocks, in particular the trimers, are akin to mechanical joints in the chain. The joints are completely reversible, in the sense that if there is no fuel for the particles there is no joint and hence no beating structures.This could point towards self assemblable micro-robots that can act in an environment-dependent manner.  

\comRS{
The inclusion of hydrodynamic interactions, which is not present in our model, is expected to change only quantitatively the results because they respect the action-reaction symmetry and by their own generate equilibrium states. The instability that generates the steady oscillations is due to the activity of the chain. In our case it comes from the phoretic interaction that breaks the action-reaction symmetry, but it could emerge in other systems with non-equilibrium interactions. A nice example of such oscillatory behavior is the magnetic chain studied in Ref.\ \cite{babel2016dynamics}, which also included hydrodynamic interactions. 
}

\comRS{To consider systems with full three dimensional motion, chains composed of more complex, three-dimensional monomers would be better candidates for steady stable oscillations. This, however, goes beyond the scope of the present paper. }

We believe self-assemblable structures pose a significant advantage over polar particles, in regard to their ease of assembly, overcoming one of the reasons of their hindered experimental exploration  \cite{Nishiguschi2017} . One can think of microfluidics systems coupled with external fields that harvest the self-assembly of smaller structures to produce colloidal chains in an assembly-line-like system. We are currently working on this idea and it will be the subject of a future publication. 

\begin{figure}[ht]
\centering
\includegraphics[width=.95\columnwidth]{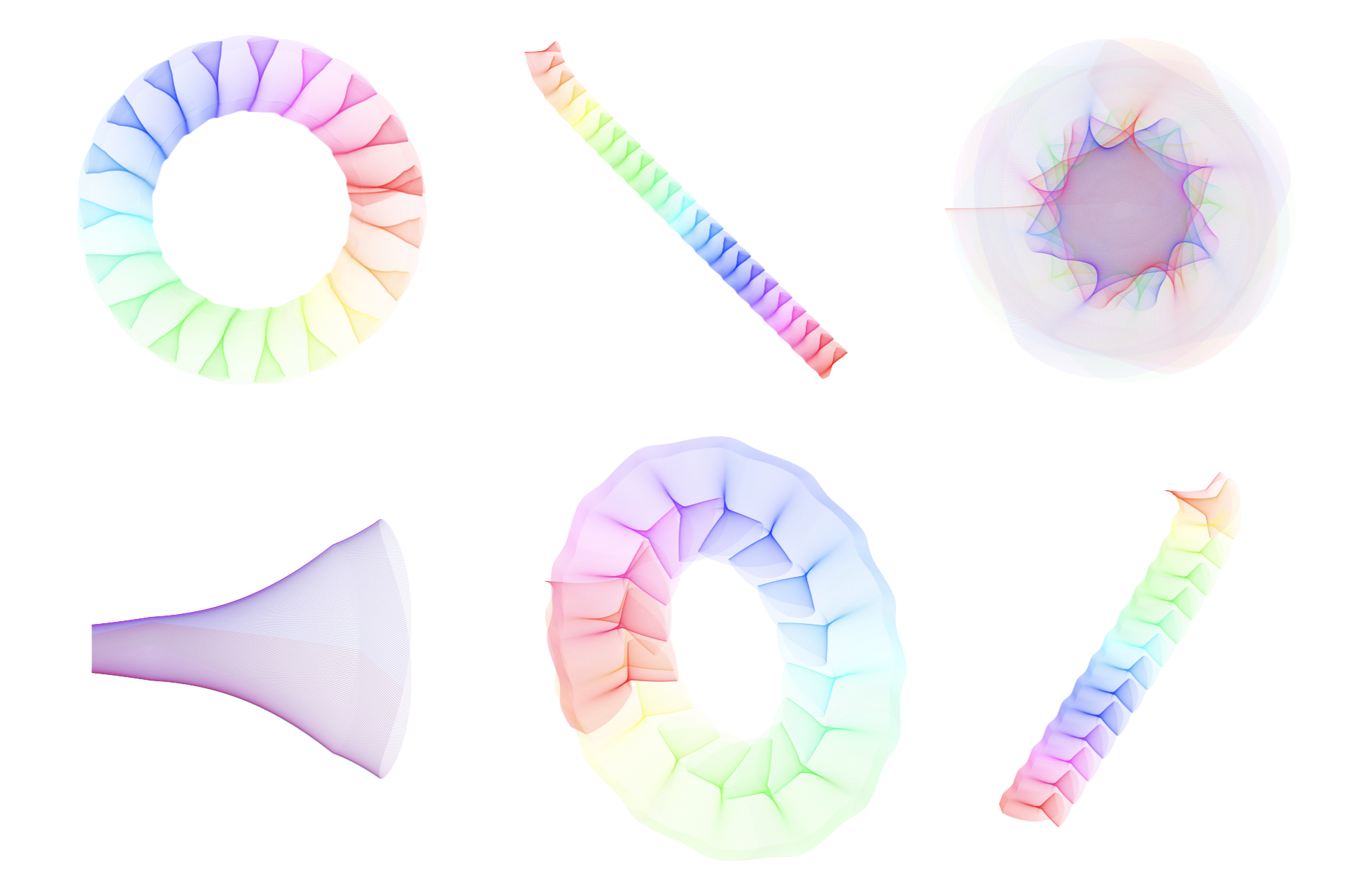}

\caption{Varieties of beating for different mobilities, charges and number of composing particles. Same color scheme as in previous figures however the time of simulation varies from 1000 to 4000 time units to show representative dynamics of the different chains. The different resulting movements go from circular trajectories, to translation in both horizontal directions, to purely transversal and space filling patterns. Upper row: left, $k=2, \alpha_A=1.5, \mu_A=0.5, M_l=M_t=0.2$; center  $k=2, \alpha_A=1.5, \mu_A=0.5, 2M_l=M_t=0.4$; right $k=2, \alpha_A=1.5, \mu_A=0.25, M_l=M_t=0.5$. Lower row: left, $k=2, \alpha_A=1.5, \mu_A=0.5, M_l=0, M_t=0.2$; center  $k=3, \alpha_A=1.5, \mu_A=0.25, M_l=0.25, M_t=0.1875$; right $k=3, \alpha_A=1.5, \mu_A=0.25, M_l=2M_t=0.25$.}
\label{fig.varities}
\end{figure}

The few examples of beating presented in this article are by far not exhaustive. Several other chain like structures have been found in the explorative stage of this research but which we decided not so to study here for the sake of simplicity, see Fig.~\ref{fig.varities} for some examples. They all share the same two basic principles: there is a head-tail  asymmetry and enough soft degrees of freedom to sustain oscillations. This can be made in many ways, and different structures present different quantitative behaviors. For example, for a beater with a certain shape and frequency, it can be designed via choosing the right size particles, chemical composition of the solute and the geometry of the chain. Furthermore, we have shown that the force exerted by these cargos is in principle tunable. This could have important implications in the design of micro robots for drug delivery, for example. Although the simulations have been done without Brownian noise, it is well known that the Hopf bifurcation and the periodic oscillations that result are robust under noise \cite{Hopf}. Hence, the mechanism we propose for generating beating filaments will continue working under experimental conditions subject to thermal noise or other source of fluctuations.

Another point worth making is the implications for our understanding of biological flagella: since the seminal work of Machin \cite{Machin1958}, the biological community has understood the movement of the sperm flagellum due to a time dependent activity of its internal motors, see Ref.~\cite{Riedel2007} for a recent example. The current results from colloidal systems (here and otherwise \cite{Nishiguschi2017,Vutukuri2017,Camalet1999}) show that this needs not to be the case: it is perfectly possible that a constant activity creates an oscillatory behavior in a filament, thus making unnecessary to assume that the motors know a priori the oscillation frequency for a sperm. This is a much simpler principle and its consequences need to be fully understood.

 \ack
 
S.G. thanks Veikko F. Geyer for the enlightening discussions on the sperm flagellum that motivated this research. This research was supported by the Fondecyt Grants No.~1140778 and 3160481, and the Millennium Nucleus ``Physics of active matter'' of the Millennium Scientific Initiative of the Ministry of Economy, Development and Tourism (Chile).

\end{document}